\documentstyle[revtex]{aps}
\begin{document}
\preprint{24 November 1993}
\draft
\begin{title}
Phason modes in spin-density wave in the presence of\\
long-range Coulomb interaction
\end{title}

\author{Attila Virosztek}
\begin{instit}
Research Institute for Solid State Physics,
H-1525 Budapest, P.O. Box 49, Hungary
\end{instit}

\author{Kazumi Maki\cite{Maki}}
\begin{instit}
Department of Physics, Hokkaido University, 060 Sapporo, Japan
\end{instit}

\receipt{}

\begin{abstract}
We study the effect of long-range Coulomb interaction on the phason in
spin-density wave (SDW) within mean field theory. In the longitudinal
limit and in the absence of SDW pinning the phason is completely
absorbed by the plasmon due to the Anderson-Higgs mechanism. In the presence
of SDW pinning or when the wave vector ${\bf q}$ is tilted from the
chain direction, though the plasmon still almost exhausts the optical
sum rule, another optical mode appears at $\omega<2\Delta(T)$, with
small optical weight. This low frequency mode below the SDW gap
may be accessible to electron energy loss spectroscopy (EELS).
\end{abstract}

\pacs{75.30.Fv, 72.15.Nj, 71.45.Gm, 78.90.+t}

\narrowtext
\section{Introduction}
It is known for some time that the long-range Coulomb interaction plays
a rather important role in phason dispersions of both charge-density
wave (CDW) \cite{Lee78,Takcdw} and spin-density wave (SDW) \cite{Taksdw}.
As a continuation of an earlier analysis of CDW \cite{VMcdw}, we study
in this paper the phason spectrum in SDW in the collisionless limit.
Unlike in CDW \cite{Pouget} the phason propagator $D_\phi$ in SDW is
not easily accessible to neutron scattering. Further, the large
frequency limit of $D_\phi$ in SDW does not allow to define the
corresponding optical weight. Therefore in the present paper we analyze
the dielectric function $\varepsilon({\bf q},\omega)$, which allows us
to define the optical weight. Moreover, not only the poles of
$\varepsilon^{-1}$ are identical to those of $D_\phi$, but also
$\varepsilon^{-1}$ is directly accessible to electron energy loss
spectroscopy (EELS).

In sharp contrast to CDW we find, that the
Anderson-Higgs mechanism \cite{AH} operates perfectly in the longitudinal
limit (i.e. when ${\bf q}$ is parallel to the chain direction) and in
the absence of SDW pinning; the phason is completely absorbed by the
plasmon and $\varepsilon$ is identical to the one in the normal state.
In the presence of SDW pinning or when ${\bf q}$ deviates from the
chain direction however, another optical mode appears below the
quasiparticle energy gap representing the plasma oscillations of the
normal (uncondensed) carriers in a dielectric environment. The optical
weight of this low frequency plasmon is always very small, but the mode
itself is well separated from the strong high frequency plasmon. In this
general circumstance the Anderson-Higgs mechanism \cite{AH} is weakly
broken.

It is worth noting that we do not find any acoustic mode in the
inverse dielectric function. In the transverse case (i.e. when
${\bf q}$ is perpendicular to the chain direction) the phason completely
decouples from the density fluctuations and therefore remains acoustic, but
because of the same decoupling none of it's features shows up in
$\varepsilon^{-1}$.

In Sec. II the general formalism for the dielectric function in SDW is
given. In Sec. III we study the pole structure in the longitudinal limit
before moving on to the general case of arbitrary direction of the wavenumber
in Sec. IV. Sec. V concludes with a brief summary of our results.

\section{Dielectric constant}
We consider the strongly anisotropic Hubbard model as introduced by
Yamaji \cite{Yamaji} supplemented by the long-range Coulomb interaction
\begin{equation}
H_C=4\pi e^2\sum_{\bf q}{1\over q^2}n_{\bf q}n_{-{\bf q}}\quad ,
\end{equation}
where $n_{\bf q}$ is the electron density operator. Within mean field
theory for the SDW the density-density correlation function \cite{VM88}
including the effect of long-range Coulomb interaction \cite{KF} is
given by
\begin{equation}
\langle [n,n]\rangle=\langle [n,n]\rangle^\prime(1+{4\pi e^2\over q^2}
\langle [n,n]\rangle^\prime)^{-1}\quad ,
\end{equation}
and
\begin{eqnarray}
\langle [n,n]\rangle^\prime=&\langle [n,n]\rangle_0+U{\langle [n,
\delta\Delta ]\rangle_0\langle [\delta\Delta,n]\rangle_0\over 1-U
\langle [\delta\Delta,\delta\Delta ]\rangle_0}\nonumber \\ =& N_0\{\langle
{\zeta^2(1-f)\over \zeta^2-\omega^2}\rangle +{\langle\zeta f\rangle^2\over
\langle (\zeta^2-\omega^2)f\rangle}\}\quad .
\end{eqnarray}
Here $\langle [n,n]\rangle^\prime$ is the correlation function obtained in
the absence of long-range Coulomb interaction,
$N_0$ is the electron density of states per spin at the Fermi energy,
$\zeta =\zeta_\parallel +\sqrt{2}\zeta_\perp\sin\varphi$, $\zeta_\parallel
=v_\parallel q_\parallel$, $\zeta_\perp =v_\perp q_\perp$, $v_\parallel$ and
$v_\perp$ are Fermi velocities parallel and perpendicular to the chain
direction respectively, and $\langle\ldots\rangle$ in the second line of
Eq.(3) means the average over $\varphi$. Finally $f(\zeta,\omega)$ is the
generalized condensate density already defined in Ref. [4b]. Expressions of
$f$ in certain limits needed in the present paper will be given later. We
note that the second term on the right hand side of Eq.(3) corresponds to
the contribution of the sliding SDW condensate.

The dielectric function defined in the usual way
\begin{equation}
\varepsilon({\bf q},\omega)=1+{4\pi e^2\over q^2}\langle [n,n]\rangle^\prime
\end{equation}
satisfies the well known relation
\begin{equation}
{\rm Im}(-\varepsilon^{-1})={4\pi e^2\over q^2}{\rm Im}\langle [n,n]
\rangle\quad .
\end{equation}
It also obeyes the following sum rule
\begin{equation}
\int_0^\infty d\omega \omega {\rm Im}(-\varepsilon^{-1})={\pi\over 2}
\omega_p^2[\cos^2\vartheta +({v_\perp\over v_\parallel})^2\sin^2\vartheta]
\quad ,
\end{equation}
where $\omega_p^2=4\pi e^2N_0v_\parallel^2=4\pi e^2n/m$ is the plasma
frquency and $\cos\vartheta=q_\parallel/q$. The total oscillator strength
depends only on the direction of ${\bf q}$.

It is worth mentioning that the phason propagator $D_\phi$ in SDW is
given by
\begin{eqnarray}
D_\phi=&(1-U\langle [\delta\Delta,\delta\Delta]\rangle_C)^{-1}\nonumber\\
=&(2\Delta)^2\{\langle (\zeta^2-\omega^2)f\rangle +{4\pi e^2\over q^2}
N_0\langle\zeta f\rangle^2\nonumber\\  &\times [1+{4\pi e^2\over q^2}N_0
\langle {\zeta^2(1-f)\over \zeta^2-\omega^2}\rangle ]^{-1}\}^{-1}\quad ,
\end{eqnarray}
where $\langle [\delta\Delta,\delta\Delta ]\rangle_C$ is the Coulomb-corrected
\cite{KF} correlation function of order parameter fluctuations. Comparing
Eqs. (2-3) and (7) it is readily seen that $D_\phi$ has the same pole
structure as $\langle [n,n]\rangle$.

\section{Longitudinal limit}
In the longitudinal limit $\zeta=\zeta_\parallel$. The correlation
function $\langle [n,n]\rangle$ is much simplified and written as
\begin{equation}
\langle [n,n]\rangle=N_0\zeta^2(\omega_p^2+\zeta^2-\omega^2)^{-1}\quad ,
\end{equation}
where we assume that the SDW condensate is free to move in the chain
direction. We see that Eq. (8) has the single pole
\begin{equation}
\omega^2=\omega_p^2+\zeta^2
\end{equation}
at the plasma frquency and $\langle [n,n]\rangle$ is the same as in the
normal state because the unpinned condensate is able to compensate
exactly for the loss of normal carriers due to the SDW transition. This
high frequency plasmon is undamped and exhausts all the available
oscillator strength Eq. (6). In other words the phason is completely
eaten by the plasmon (Anderson-Higgs mechanism \cite{AH}).

This situation is slightly modified in the presence of SDW pinning. For
simplicity let us assume that the pinning is so strong that it suppresses
completely the fluctuation of the order parameter $\delta\Delta$ induced
by the density fluctuation $n_{\bf q}$. This amounts to neglecting the
second term (the contribution of the condensate) in Eq. (3). Then we obtain
\begin{equation}
\langle [n,n]\rangle_{pinned}=N_0\zeta^2(1-f)[\omega_p^2(1-f)+\zeta^2-
\omega^2]^{-1}\quad .
\end{equation}
Due to the imaginary part of $f$ a weak single particle continuum appears
in ${\rm Im}(-\varepsilon^{-1})$ for $0<\omega<|\zeta|$ and for
$\omega>\sqrt{(2\Delta)^2+\zeta^2}$ with square root edge at $\omega=
\sqrt{(2\Delta)^2+\zeta^2}$ and with $\omega^{-4}$ decay for $\omega
\rightarrow\infty$. The high frequency plasmon is modified to
\begin{equation}
\omega^2=\omega_p^2+(2\Delta)^2[\ln({\omega_p\over\Delta})-i{\pi\over 2}]
+\zeta^2\quad ,
\end{equation}
where we used the asymptotic expansion
\begin{equation}
f(\omega\gg 2\Delta)\approx -({2\Delta\over\omega})^2[\ln({\omega\over
\Delta})-i{\pi\over 2}]\quad .
\end{equation}
This plasmon stiffens and broadens somewhat due to the pinning, but
remains relatively sharp and still exhausts almost all the oscillator
strength.

There appears however another optical mode below the SDW gap $2\Delta$.
This low frequency mode is undamped because it can not decay into
single particle excitations, and it's frequency is given by
\begin{equation}
f(\omega)=1\quad ,
\end{equation}
where we made use of the relation $2\Delta\ll\omega_p$. At low temperature
($T\ll T_c$) this mode is well below the gap, and we can use the
low frequency expansion
\begin{equation}
f(\omega)=f_d+{2\over 3}({\omega\over 2\Delta})^2(1-g_d)+\ldots\quad ,
\end{equation}
where
\begin{equation}
f_d=1-2\int_0^\infty d\varphi \cosh^{-2}\varphi [1+\exp(\beta\Delta
\cosh\varphi)]^{-1}
\end{equation}
is the condensate density in the dynamic limit,
\begin{equation}
g_d=3\int_0^\infty d\varphi \cosh^{-4}\varphi [1+\exp(\beta\Delta
\cosh\varphi)]^{-1}\quad ,
\end{equation}
and $\beta=1/T$. The low frequency mode is then given by
\begin{equation}
\omega^2={3\over 2}(2\Delta)^2{1-f_d\over 1-g_d}\approx {3\over 2}
(2\Delta)^2\sqrt{2\pi T\over\Delta}e^{-\Delta/T}\quad ,
\end{equation}
and the corresponding optical weight is evaluated as
\begin{equation}
{\pi\over 2}\omega_p^2({3\over 2})^2({2\Delta\over\omega_p})^4
{1-f_d\over (1-g_d)^2}\quad .
\end{equation}
This mode is clearly the plasma oscillation due to the normal carriers
embedded in a dielectric with a $2\Delta$ gap produced by the pinned SDW
condensate. The optical weight is very small ($\sim (2\Delta/\omega_p)^4$)
and vanishes exponentially for $T\rightarrow 0$ just like the frequency of the
mode itself.

If the temperature approaches $T_c$, the frequency of the lower plasmon
moves close to the SDW gap, and Eq. (14) is solved using the limiting
expression
\begin{equation}
f(\omega\rightarrow 2\Delta)={\pi\over 2}{\tanh(\Delta/2T)\over
\sqrt{1-(\omega/2\Delta)^2}}-f_s\quad ,
\end{equation}
where
\begin{equation}
f_s=\pi\Delta^2 T\sum_\omega(\omega^2+\Delta^2)^{-3/2}
\end{equation}
is the condensate density in the static limit. The frequency of the mode
is given by
\begin{equation}
\omega^2=(2\Delta)^2[1-({\pi\Delta\over 4T})^2]\quad ,
\end{equation}
and the corresponding relative optical weight is
\begin{equation}
2({2\Delta\over\omega_p})^4({\pi\Delta\over 4T})^2\quad ,
\end{equation}
which vanishes like $(T_c-T)^3$ as $T$ approaches $T_c$. Clearly, the
optical weight of this midgap plasmon is maximum at intermediate temperature,
when the frequency of the mode is well separated from the gap edge, but still
of the order of the gap value.

\section{Optical modes for arbitrary\\
direction of the wavenumber}
When ${\bf q}$ is not parallel to the chain direction, the angular
averages in Eq. (3) are not easy to perform in general. However, since
we want to follow the evolution of the optical modes described above
in the previous Section, we can study the $q\rightarrow 0$ limit, keeping
only the information about the direction of ${\bf q}$.

If the SDW condensate is pinned, we find that the results of Sec. III
for the pinned case still apply with only one modification. The plasma
frquency $\omega_p^2$ should be replaced by
\begin{equation}
\omega_p^2(\vartheta)=\omega_p^2[\cos^2\vartheta+({v_\perp\over
v_\parallel})^2\sin^2\vartheta]\quad ,
\end{equation}
which determines the direction dependence of the higher plasma frequency.
$\omega_p(\vartheta)$ is still
large compared to $2\Delta$ even in the transverse limit $\cos\vartheta=0$,
therefore Eqs. (13), (17), and (21) describing the frequency of the lower
plasmon are valid in this case as well with no direction dependence. It
can be seen from Eqs. (18) and (22) however, that the optical weight of the
low frequency plasmon will increase by a factor of $(v_\parallel/v_\perp)^2
\approx 10^2$ while $\vartheta$ increases from zero to $\pi/2$. Therefore
the lower plasmon is easiest to detect in the transverse limit.

If the SDW is unpinned, Eq. (3) reduces to
\begin{equation}
\langle [n,n]\rangle^\prime=-{N_0\over\omega^2}\{\zeta_\parallel^2+
\zeta_\perp^2[1-f(\omega)]\}
\end{equation}
in the long wavelength limit. The corresponding dielectric function is
\begin{equation}
\varepsilon(\vartheta,\omega)=1-{\omega_p^2\over\omega^2}\{\cos^2\vartheta+
({v_\perp\over v_\parallel})^2\sin^2\vartheta[1-f(\omega)]\}\quad .
\end{equation}
The high frequency plasmon mode is again given by Eq. (23) and exhausts
almost all the oscillator strength. We note that although the corresponding
peak in ${\rm Im}(-\varepsilon^{-1})$ is somewhat broadened in general, it
becomes undamped in the longitudinal limit as it should according to Eq. (8).

The low frequency plasmon mode is obtained from Eq. (25) as the solution to
\begin{equation}
f(\omega)=1+(\zeta_\parallel/\zeta_\perp)^2\quad .
\end{equation}
Note, that in the transverse limit ($\zeta_\parallel=0$) Eq. (26) is identical
to Eq. (13) describing this mode in the pinned case. Otherwise the mode is
always higher in the unpinned case. If $\zeta_\parallel\ll
\zeta_\perp$ and $T\ll T_c$ then the mode is well below the gap, and
Eqs. (14-16) are used to evaluate it's frequency as
\begin{equation}
\omega^2={3\over 2}(2\Delta)^2[1-f_d+(\zeta_\parallel/\zeta_\perp)^2]
(1-g_d)^{-1}\quad .
\end{equation}
The relative weight of this mode is
\begin{equation}
({3\over 2})^2({2\Delta\over\omega_p(\vartheta)})^4[1-f_d+(\zeta_\parallel/
\zeta_\perp)^2](1-g_d)^{-2}\quad ,
\end{equation}
which is still small, but neither the frequency of the mode, nor it's
weight freezes out for $T\rightarrow 0$ if $\zeta_\parallel\neq 0$.

In the other limit, i.e. if $\zeta_\parallel\gg\zeta_\perp$, or $T\rightarrow
T_c$, we use Eqs. (19-20) in order to solve Eq. (26) as
\begin{equation}
\omega^2=(2\Delta)^2\{1-[{\pi\over 2}{\tanh(\Delta/2T)\over 1+f_s+
(\zeta_\parallel/\zeta_\perp)^2}]^2\}\quad .
\end{equation}
The corresponding relative optical weight is now given by
\begin{eqnarray}
{\pi^2\over 2}&({2\Delta v_\parallel\over\omega_p v_\perp\sin\vartheta})^4
\tanh^2({\Delta\over 2T})[1+(\zeta_\parallel/\zeta_\perp)^2]^{-1}
\nonumber \\ & \times [1+f_s+(\zeta_\parallel/\zeta_\perp)^2]^{-3}\quad ,
\end{eqnarray}
which is small and vanishes completely in the longitudinal limit where
only the high frequency plasmon is present as in Eq. (8).

It is to be noted that our results for the unpinned and pinned cases
refer to two extreme situations, namely the pinning strength being zero and
infinite respectively. Incorporating finite pinning strength, or in other
words discussing the situation when the fluctuations of the order
parameter is not completely suppressed,
is beyond the scope of this communication. Clearly,
further work in this direction is desirable. However, in the experimentally
interesting transverse geometry, where the predicted low frequency plasmon
mode is most likely to be observable, the actual pinning strength does not
play a crucial role, since the characteristics of this mode are identical
for the pinned and unpinned cases in the transverse limit.

\section{Concluding Remarks}
We have extended an earlier analysis of the effect of long-range Coulomb
interaction in CDW to the phason in SDW. In contrast to CDW we find that
the Coulomb interaction has pervasive effect on the phason. Except in the
transverse limit where the Coulomb interaction does not play a role, no
acoustic mode is present (Anderson-Higgs mechanism). This is due to the
fact that in SDW there is no effective mass enhancement and the velocity of
the phason would be that of the Fermi velocity. The normal carriers,
having at most the same velocity, are unable to screen the fast phason and
only optical modes survive. The optical sum rule for the inverse dielectric
function is almost completely exhausted by the high frequency plasmon
which is present already in the normal state. Below $T_c$ however, there
is a small contribution of another plasmon mode at a frequency
$\omega<2\Delta(T)$ due to the uncondensed carriers
in a dielectric environment, which could in principle be accessible
to EELS, most likely close to the transverse limit at intermediate
temperatures.

\acknowledgments
This publication is sponsored by the U.S.-Hungarian Science and
Technology Joint Fund in cooperation with the National Science Foundation
and the Hungarian Academy of Sciences under Project No. 264/92a, which
enabled one of us (A. V.) to enjoy the hospitality of USC. K. M.
acknowledges greatfully a Japanese Society of Promotion of Science
Fellowship for giving him opportunity to work at Hokkaido University. This
work is also supported in part by the National Science Foundation under
Grant No. DMR92-18317 and by the Hungarian National Research Fund under
Grant No. OTKA2944 and T4473.


\begin{references}
\bibitem[*]{Maki}On sabbatical leave from Department of Physics and
Astronomy, University of Southern California, Los Angeles,
California 90089-0484.
\bibitem{Lee78} P. A. Lee and H. Fukuyama, Phys. Rev. B {\bf 17},
542 (1978).
\bibitem{Takcdw} Y. Nakane and S. Takada, J. Phys. Soc. Jpn. {\bf 54},
977 (1985); K. Y. Wong and S. Takada, Phys. Rev. B {\bf 36}, 5476 (1987).
\bibitem{Taksdw} Y. Nakane, H. Miyazawa, and S. Takada, J. Phys. Soc. Jpn.
{\bf 57}, 3424 (1988).
\bibitem{VMcdw} A. Virosztek and K. Maki, Phys. Rev. Lett. {\bf 69}, 3265
(1992); A. Virosztek and K. Maki, Phys. Rev. B {\bf 48}, 1368 (1993).
\bibitem{Pouget} B. Hennion, J. P. Pouget, and M. Sato, Phys. Rev. Lett.
{\bf 68}, 2374 (1992).
\bibitem{AH} P. W. Anderson, Phys. Rev. {\bf 112}, 1900 (1958);
G. Rickayzen, Phys. Rev. {\bf 115}, 795 (1959).
\bibitem{Yamaji} K. Yamaji, J. Phys. Soc. Jpn. {\bf 51}, 2787 (1982);
{\bf 52}, 1361 (1983).
\bibitem{VM88} A. Virosztek and K. Maki, Phys. Rev. B {\bf 37}, 2028 (1988).
\bibitem{KF} L. P. Kadanoff and I. I. Falko, Phys. Rev. {\bf 136}, 1170
(1964).
\end{references}
\end{document}